\documentclass{elsart}
\usepackage{graphics,harvard,amssymb}
\journal{New Astronomy}

\makeatletter
\def\elsartstyle{%
	\def\normalsize{\@setfontsize\normalsize\@xiipt{14.5}}
	\def\small{\@setfontsize\small\@xipt{13.6}}
	\let\footnotesize=\small
	\def\large{\@setfontsize\large\@xivpt{18}}
	\def\Large{\@setfontsize\Large\@xviipt{22}}
	\skip\@mpfootins = 18\p@ \@plus 2\p@
	\normalsize
}
\makeatother
\def\degr{\hbox{$^\circ$}}

\def\bibcode#1{(\texttt{#1})}
\def\astrobj#1{#1}
\def\url#1{{\ttfamily\def\/{/\discretionary{}{}{}}#1}}

\begin{document}

\begin{frontmatter}
\title{Solar-like oscillations in the F9\,V $\beta$~Virginis\thanksref{titleobs}}
\thanks[titleobs]{Based on observations obtained at the 1.2-m
Swiss Euler telescope at La Silla (ESO, Chile)}
      
\author[Geneve]{F. Carrier\thanksref{email}},
\author[Geneve]{P. Eggenberger},
\author[Geneve]{A. D'Alessandro},
\author[Geneve]{L. Weber}
\address[Geneve]{Observatoire de Gen\`eve, 51 Ch. des Maillettes, CH--1290 Sauverny, Suisse}

\thanks[email]{E-mail: Fabien.Carrier@obs.unige.ch}

\begin{abstract}
This paper presents the analysis of Doppler $p$-modes of the F9\,V star
\astrobj{$\beta$~Virginis} obtained with the spectrograph \textsc{Coralie} in March 2003.
Eleven nights of observations have made it possible to collect 1293 radial velocity
measurements with a standard deviation of about 2.2\,m\,s$^{-1}$. The power spectrum of the
high precision velocity time series clearly presents several identifiable peaks between 0.7 and
2.4\,mHz showing regularity with a large and small spacings of $\Delta\nu$\,=\,72.1\,$\mu$Hz and
$\delta\nu_{02}$\,=\,6.3\,$\mu$Hz respectively. Thirty-one individual modes have been
identified with amplitudes in the range 23 to 46\,cm\,s$^{-1}$, i.e. with a signal to noise between
3 and 6.
\end{abstract}

\end{frontmatter}

\section{Introduction}
\label{intro}
The lack of observational constraints leads to serious uncertainties in the
modeling of stellar interiors.
The measurement and characterization of oscillation modes
is an ideal tool to test models of stellar inner structure and
theories of stellar evolution. Indeed,
the five-minute oscillations in the Sun have led to a wealth of information
about the solar interior. These results stimulated various attempts to detect
a similar signal on other solar-like stars.
Solar--like oscillation modes generate periodic motions of the stellar
surface with periods in the range of 3--60 minutes but with extremely small
amplitudes. Essentially, two methods exist to detect such a motion:
photometry and Doppler spectroscopy. In photometry, the oscillation 
amplitudes of solar--like stars
are within 2--30~ppm, while they are in the range of
10--150~cm\,s$^{-1}$ in radial velocity measurements.
Photometric measurements made from the ground are strongly limited by
scintillation noise. To reach the needed accuracy requires
observations made from space.
In contrast, Doppler ground--based measurements have recently shown their
ability to detect oscillation modes in solar--like stars. 
These past years, the stabilized spectrographs developed
for extra-solar planet detection achieved accuracies needed for solar-like 
oscillation detection by means of radial velocity measurements
\cite{carrier1}. 

A primary target for the search for $p$-mode oscillations is the bright dwarf
F9 \astrobj{$\beta$~Virginis} (HR~4540, HD~102870, m$_V$\,=\,3.61). 
Scaling from the solar case, $p$-modes are
expected near $\nu_{\rm max}$\,=\,1.4\,mHz with a large frequency spacing of about 
$\Delta\nu_{0}$\,=\,71\,$\mu$Hz,
and a maximal amplitude of A$_{\rm osc}$\,=\,65\,cm\,s$^{-1}$ 
\cite{kb95}.

In this paper, we report Doppler observations 
of \astrobj{$\beta$~Vir} made with the \textsc{Coralie} spectrograph, well known for 
the characterization of p--modes on the $\alpha$ Cen system \cite{bc02,cb03}.
These measurements enable the identification of thirty-one individual mode frequencies.

\section{Observations and data reduction}
\label{datareduc}
\astrobj{$\beta$~Vir} was observed over a campaign of eleven nights (2003 February 28 - March 10)
with \textsc{Coralie}, the high-resolution fiber-fed echelle spectrograph mounted on the
1.2-m Swiss telescope at La Silla (ESO, Chile). During the stellar exposures, the spectrum of a
thorium lamp carried by a second fiber is simultaneously recorded in order to
monitor the spectrograph's stability and thus to obtain high-precision
velocity measurements.  The wavelength coverage of the
spectra is 3875-6820\,\AA , recorded on 68 orders. 
The dead-time between two exposures was improved from the usual 125\,s to 85\,s,
by archiving the image during the following exposure.
Exposure times of 120\,s, thus cycles of 205\,s,
allowed us to obtain 1293 spectra, with a typical signal-to-noise ratio (S/N) 
in the range of 100--140 at 550\,nm.
\begin{figure}
\resizebox{\hsize}{!}{\includegraphics{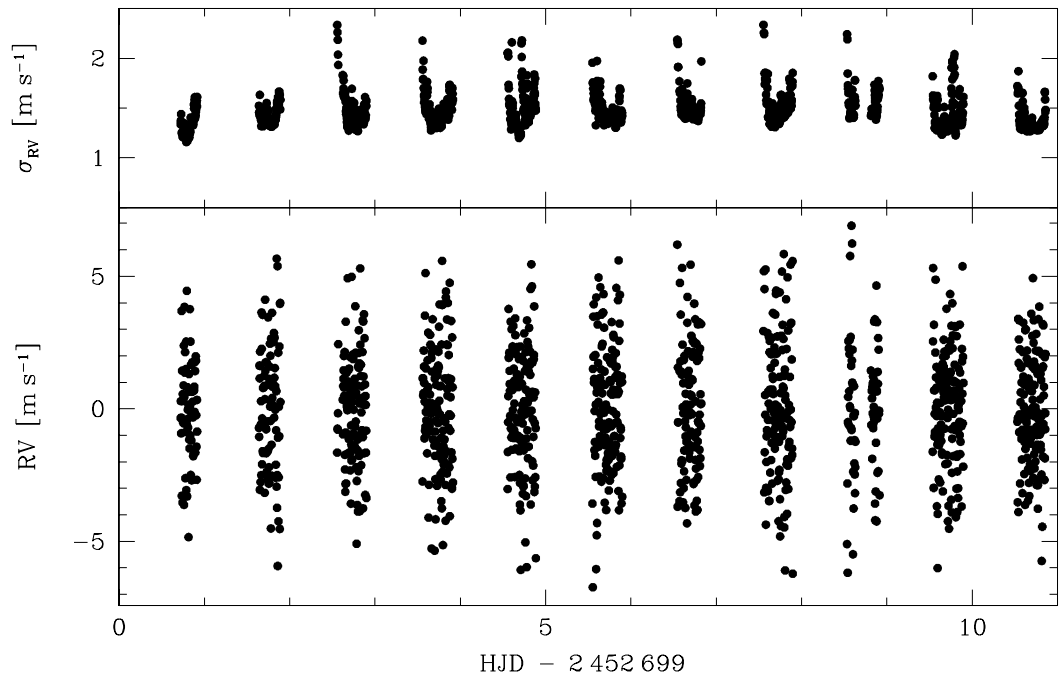}}
\caption[]{Radial-velocity measurements of \astrobj{$\beta$ Vir}. The dispersion
(which includes the noise and the oscillations) reaches 2.24\,m\,s$^{-1}$. 
The upper pannel represents the photon noise uncertainties.}
\label{fig:rv}
\end{figure}

\begin{table}
\caption{Distribution and dispersion of Doppler measurements}
\begin{center}
\begin{tabular}{clll}
\hline
\hline
Date & Nb spectra & Nb hours & $\sigma$ (m\,s$^{-1}$) \\ \hline
2003/02/28 & 69  & 4.67 & 1.91 \\
2003/03/01 & 91  & 6.06 & 2.35 \\
2003/03/02 & 118 & 8.21 & 2.01 \\
2003/03/03 & 137 & 8.55 & 2.25 \\
2003/03/04 & 135 & 7.95 & 2.22 \\
2003/03/05 & 142 & 8.37 & 2.33 \\
2003/03/06 & 111 & 6.75 & 2.29 \\
2003/03/07 & 132 & 8.24 & 2.60 \\
2003/03/08 & 76  & 9.13 & 2.54 \\
2003/03/09 & 146 & 8.64 & 2.12 \\
2003/03/10 & 136 & 7.92 & 2.03 \\ \hline
\label{tab:journal}
\end{tabular}
\end{center}
\end{table}

Radial velocities
are computed for each night relative to the highest
S/N spectrum obtained in the middle of the night by the use of the 
optimum-weight procedure \cite{co85,carrier2}.
This method
requires a Doppler shift that remains small compared to the line-width (smaller
than 100\,m\,s$^{-1}$). Since the Earth's motion
can introduce a Doppler shift larger than 700\,m\,s$^{-1}$ during a whole night,
each
spectrum is first corrected for the Earth's motion before deriving the radial
velocities. Subsequently, the mean for each night is subtracted.
The rms scatter of the time series is 2.24\,m\,s$^{-1}$ (see Fig.~\ref{fig:rv} and Table~\ref{tab:journal}), 
which can be directly
compared with the fundamental uncertainty due to photon noise.
The uncertainties coming from the thorium spectrum used in the instrumental tracking are
quite stable with a value of about 0.34\,m\,s$^{-1}$.
The quadratic sum of the stellar spectrum photon noise and this instrumental photon noise varies
between 1.2 and 2.3\,m\,s$^{-1}$.

\section{Power spectrum analysis}
\subsection{\textsc{Coralie} measurements}
\begin{figure}
\resizebox{\hsize}{!}{\includegraphics{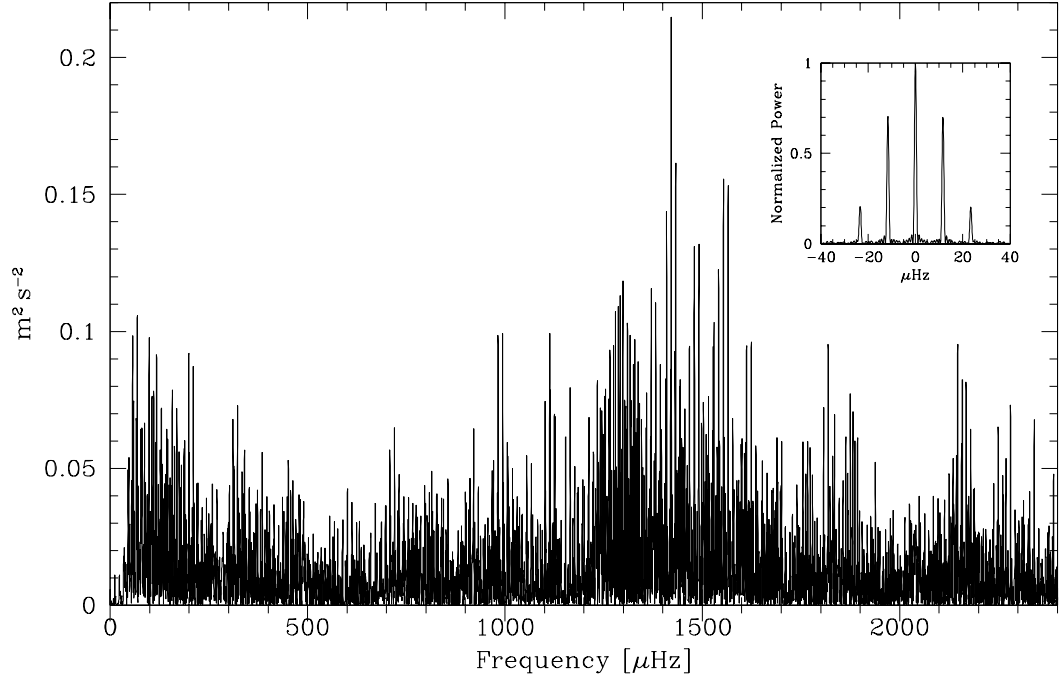}}
\caption{Power spectrum of the \textsc{Coralie} radial velocity measurements of \astrobj{$\beta$ Vir}.
The window function is shown in the inset.}
\label{fig:tf}
\end{figure}

In order to compute the power spectrum of the velocity time series, we use
the Lomb-Scargle modified algorithm \cite{lomb,scargle}
with a weight being assigned to each
point according its uncertainty estimate. The time scale gives a formal resolution
of 1.14\,$\mu$Hz. The Nyquist frequency (cut-off) is determined by the measurement cycle of 205\,s and has a value
of $\nu_N$\,$\simeq$\,2.4\,mHz. The resulting periodogram, shown in Fig.~\ref{fig:tf},
exhibits a series of peaks centred at 1.4\,mHz, exactly where
the solar-like oscillations for this star are expected.
Typically for such a power spectrum, the noise has two components:
\begin{itemize}
\item Towards the lowest frequencies, the power should scale inversely with
frequency squared, as expected for instrumental instabilities. 
However, the computation of the radial velocities introduces a high pass filter. Indeed, the radial velocities were computed relative to one
reference for each night and the average radial velocities of the night fixed to zero (see Sect.~\ref{datareduc}). This results in an 
attenuation of the very low frequencies which can be seen on Fig.~\ref{fig:tf}.
\item At high frequencies it is flat, indicative
of the Poisson statistics of photon noise. 
As $p$-modes are present until a frequency of 2.4\,mHz (see Sect.~\ref{harps}), we have to compute this noise at lower
frequency than the oscillations. The noise was determinated per interval: it is decreasing until 530\,$\mu$Hz (instrumental instabilities)
and is increasing
again from 690\,$\mu$Hz due to the oscillation modes. We thus calculated the flat noise in the interval 530\,--\,690\,$\mu$Hz which 
seems constant with a value of 0.0074\,m$^2$\,s$^{-2}$, namely 7.6\,cm\,s$^{-1}$ in amplitude.
With 1293 measurements, this high frequency noise 
corresponds to a velocity accuracy of $\sigma_{RV}\,=\,\sqrt{N \sigma_{\mathrm{pow}} /4 }\,=\,1.55$~m\,s$^{-1}$. 
\end{itemize}
The power spectrum presents an excess in the range 0.7--2.4~mHz (see also Sect.~\ref{harps}). 
Note that the filtering induced by the radial velocities computation does not influence the frequency of the peaks in the 
range 0.7--2.4~mHz, but could slightly change their amplitudes.
The amplitude of the strongest peaks reaches 46~cm\,s$^{-1}$, corresponding to a signal to noise of 6 (in the amplitude spectrum).
\subsection{\textsc{Harps} measurements}
\label{harps}
\begin{figure}
\resizebox{\hsize}{!}{\includegraphics{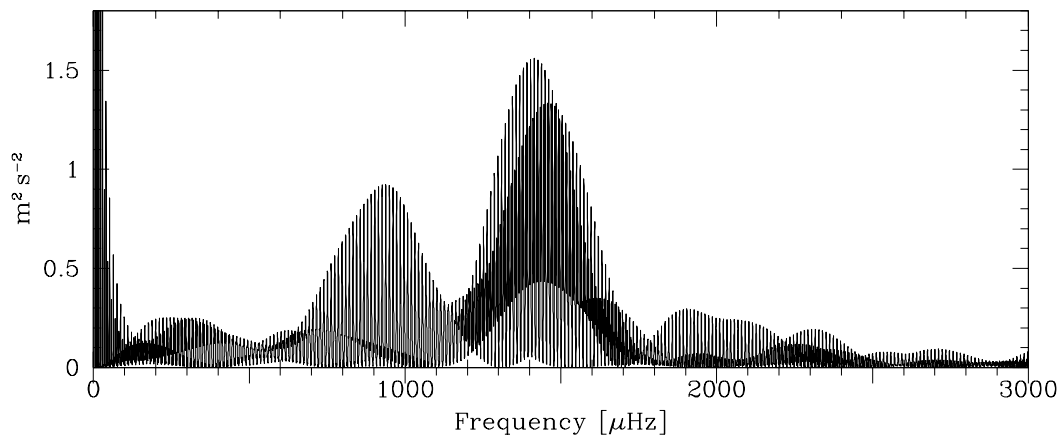}}
\caption{Power spectrum of the \textsc{Harps} radial velocity measurements of \astrobj{$\beta$ Vir}
(see Sect~\ref{harps}).
The daily aliases are too high to allow any mode determination, however, we can see that $p$-modes
are present in the range 0.7\,--\,2.4\,mHz.}
\label{fig:tfharps}
\end{figure}
Some measurements were obtained with the spectrograph \textsc{Harps} installed on the 3.6-m telescope at La Silla
Observatory (ESO, Chile) \cite{pepe} in July 2004 in order to 
better determine the range of $p$-mode frequencies. 171 measurements were collected during 5 nights and the resulting
power spectrum is shown in Fig.~\ref{fig:tfharps}. Due to observations of only 40\,minutes per night, these measurements
cannot allow us to determine mode frequencies: the five first daily aliases (until 5\,$\times$\,11.57\,$\mu$Hz on each side) have an
amplitude greater than 90\% of the main mode: it is impossible to safely choose the right frequencies. 
However, due to a quicker observation cycle and thus to a higher Nyquist frequency, we can remark that oscillation
modes are present in the range 0.7\,--\,2.4\,mHz.

\subsection{Search for a comb-like pattern}

\begin{figure}
\resizebox{\hsize}{!}{\includegraphics{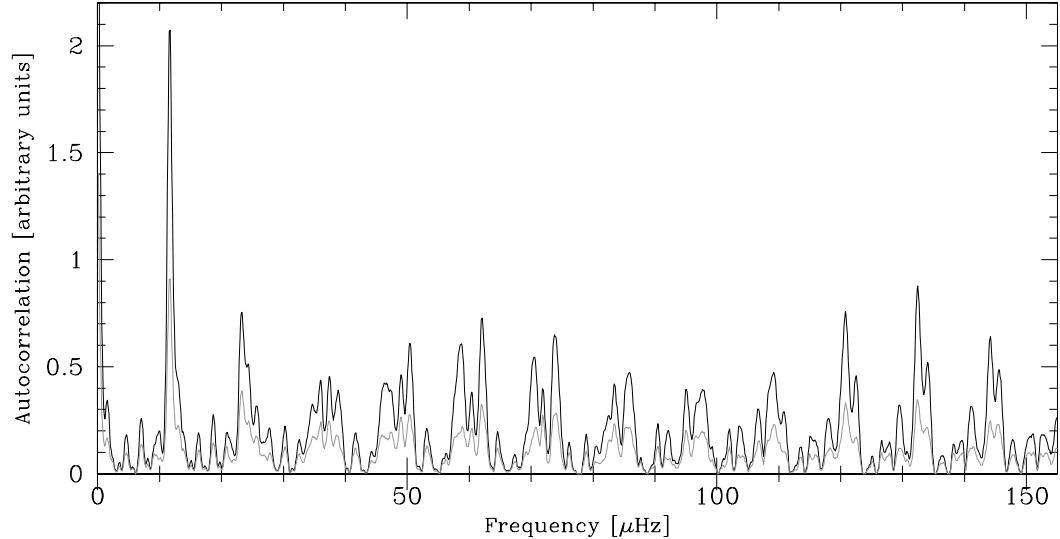}}
\caption{Autocorrelation of the power spectrum with a threshold of 0.07\,m$^2$\,s$^{-1}$. The gray line
corresponds to the same autocorrelation but all peaks greater than the threshold have the 
same amplitude (not to favour only the highest peaks). 
The large spacing is estimated to be about 72\,$\mu$Hz.}
\label{fig:ac}
\end{figure}
In solar-like stars, p-mode oscillations of low-degree are expected to produce a
characteristic comb-like structure in the power spectrum with mode
frequencies
$\nu_{n,l}$ reasonably well approximated by the asymptotic
relation \cite{tassoul80}:
\begin{eqnarray}
\label{eq1}
\nu_{n,l} & \approx &
\Delta\nu(n+\frac{l}{2}+\epsilon)-l(l+1) D_{0}\;.
\end{eqnarray}
Here, $D_0$, which is equal to $\frac{1}{6} \delta\nu_{02}$ if the 
asymptotic relation holds exactly, and $\epsilon$ are sensitive to the sound speed near the core and to
the surface layers respectively. 
The quantum numbers $n$ and $l$ correspond to the radial
order and the angular degree of the modes, and $\Delta\nu$ and
$\delta\nu_{02}$
to the large and small spacings.
To fit to this relation, an autocorrelation of the power spectrum
is calculated and presented in Fig.~\ref{fig:ac}. The threshold of the power spectrum (all frequencies
with a smaller amplitude than this value has its amplitude fixed to zero) is chosen to 0.07\,m$^2$\,s$^{-1}$
corresponding to a signal to noise of 3.5 in amplitude, this process can diminish the noise.
Each peak of the autocorrelation corresponds to a structure present in the power spectrum.
The two strong peaks at low frequency near 11.6 and 23.1\,$\mu$Hz
correspond to the daily aliases. The strongest other peaks are situated near 60\,$\pm$\,11.57\,$\mu$Hz
and 130\,$\pm$\,11.57\,$\mu$Hz which correspond respectively to one and two times the large spacing.
We can thus deduce that the large spacing is near 72\,$\mu$Hz, as peaks at 2 times a separation of 60 or 80\,$\mu$Hz
are not present, this is in accordance with theoretical predictions
(see Sect.~\ref{intro}). Three peaks coexist near 72\,$\mu$Hz, namely 70.5, 72 and 74\,$\mu$Hz, but
the most probable large spacing
is situated at 72\,$\mu$Hz: a high peak has the value of 2 times this spacing whereas only a low peak is present
near 141\,$\mu$Hz, and no peak at all near 148\,$\mu$Hz. This large spacing will be in any case refined 
during the following step.

\subsection{Mode identification}
\label{mi}
\begin{table}
\caption[]{Identification of extracted frequencies. The number of peaks due to noise is in agreement with the simulations
described in Sect.~\ref{mi}, which predict 4.9\,$\pm$\,4.3 noise peaks.}
\begin{center}
\begin{tabular}{rcc}
\hline
\hline
\multicolumn{1}{c}{Frequency} & Mode ID  & S/N \\
\multicolumn{1}{c}{$[\mu$Hz$]$} &           &     \\
\hline
$720.1 + 11.6 = 731.7$ &  $\ell = 1$   & 3.2 \\
$814.6 - 11.6 = 803.0$ &  $\ell = 1$   & 3.0\\
$920.9 - 11.6 = 909.3$ &  $\ell = 0$   & 3.3\\
982.6  &  $\ell = 0$   & 4.2\\
$1113.3 + 11.6 = 1124.9$  &   $\ell = 0$   & 4.1\\
$1165.3 - 11.6 = 1153.7$ &   $\ell = 1$   & 3.8\\
$1212.6 + 11.6 = 1224.2$ &   $\ell = 1$   & 3.1\\
1258.7 &  $\ell = 2$   &  3.3\\
1266.1 &  $\ell = 0$   &  4.1\\
1291.7 &   noise   & 4.1\\ 
1298.7 &   $\ell = 1$   & 4.6\\
$1329.1 + 11.6 = 1340.7$ & $\ell = 0$   & 3.9\\
$1382.2 - 11.6 = 1370.6$ &   $\ell = 1$   & 4.6\\
$1391.8 + 11.6 = 1403.2$ & $\ell = 2$ & 3.0 \\
$1421.2 - 11.6 = 1409.6$ &   $\ell = 0$   & 6.1\\
$1430.9 + 11.6 = 1442.5$ &   $\ell = 1$   & 3.9\\
1447.9 &   noise   & 3.3\\  
1478.5 &   $\ell = 2$   & 4.2\\
1479.9 &   $\ell = 2$   & 4.5\\
$1527.8 - 11.6 = 1516.2$ &   $\ell = 1$   & 4.5\\
1553.6 &  $\ell = 2$   &  5.2\\
$1598.7 - 11.6 = 1587.1$ &  $\ell = 1$   &  3.3\\
1624.1 &  $\ell = 2$   &  4.0\\
1697.8 &  $\ell = 2$   &  3.0\\
$1691.1 + 11.6 = 1702.7$ &  $\ell = 0$   &  3.1\\
$1754.9 + 11.6 = 1766.5$&  $\ell = 2$   &  3.1\\
$1818.9 - 11.6 = 1807.3$&  $\ell = 1$   &  3.9\\
$1835.0 + 11.6 = 1846.6$&  $\ell = 0$   &  3.5\\
1875.2 &  $\ell = 1$   &  3.6\\
$2147.0 - 11.6 = 2135.4$&  $\ell = 0$   &  4.1\\
2168.4 &  $\ell = 1$   &  3.2\\
$2249.8 - 11.6 = 2238.2$&  $\ell = 1$   &  3.2\\
$2340.9 + 11.6 = 2352.5$&  $\ell = 0$   &  3.4\\
\hline
\end{tabular}
\end{center}
\label{tab:identif}
\end{table}

\begin{figure}[thb]
\resizebox{\hsize}{!}{\includegraphics{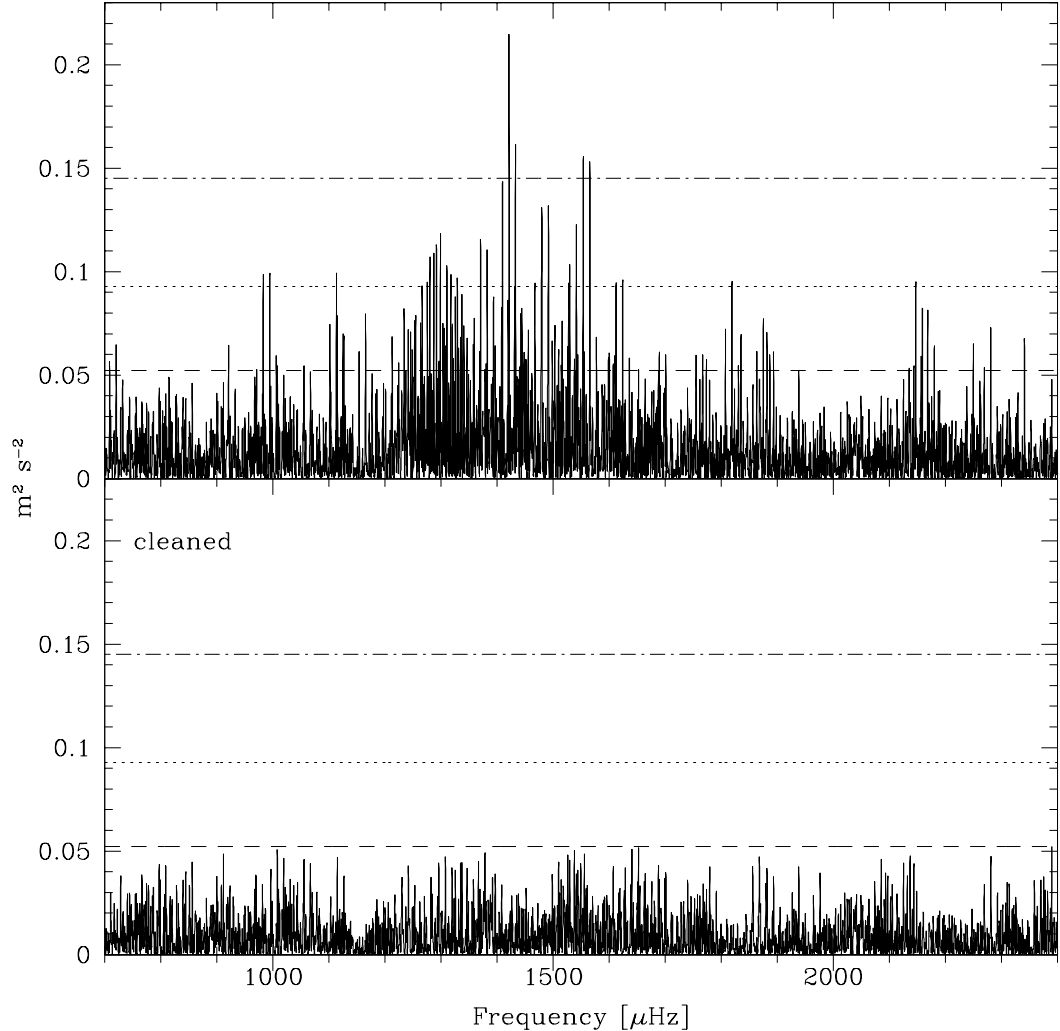}}
\caption[]{{\bf Top:} Original power spectrum of \astrobj{$\beta$ Vir}. {\bf Bottom:} Cleaned power spectrum: all peaks listed
in Table~\ref{tab:identif} have been removed. The dot-dashed, dotted and dashed lines indicate an amplitude of 5\,$\sigma$, 
4\,$\sigma$ and 3\,$\sigma$, respectively. Numerous peaks are still present below 3\,$\sigma$, 
since no peaks have been cleaned below this threshold. These peaks can be due to $p$--mode oscillations and noise or have 
artificially been
added by the extraction algorithm due to the finite lifetimes of the modes} 
\label{clean}
\end{figure}

\begin{figure*}[thb]
\begin{center}
\resizebox{\hsize}{!}{\includegraphics{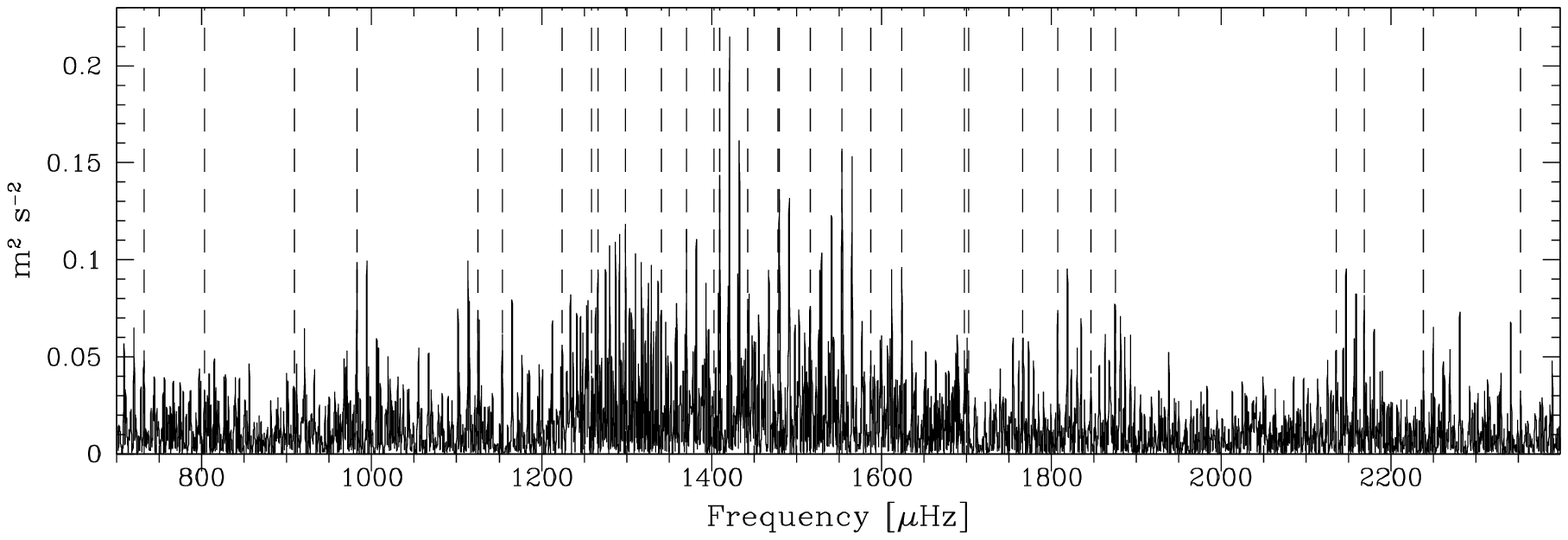}}
\caption[]{Power spectrum of \astrobj{$\beta$ Vir} with the thirty-one extracted frequencies indicated by dashed lines.
The identification of each extracted frequency is given in Table~\ref{tab:freq}.}
\label{tfiden}
\end{center}
\end{figure*}

The frequencies were extracted using an iterative algorithm, which identifies the highest peak between
700 and 2400\,$\mu$Hz and subtracts
it from the time series. 
Note that because of the stochastic nature of solar--like oscillations, a timestring
of radial velocities cannot be expected to be perfectly reproduced by a sum of sinusoidal terms. Therefore, using
an iterative clean algorithm to extract the frequencies can add additional peaks with small amplitudes due to the finite
lifetimes of the modes that we do not know. Nevertheless, the iterative algorithm ensures that one peak and its aliases
with an amplitude above a given threshold is only extracted once.
To avoid extracting artificial peaks with small amplitudes added by the iterative
algorithm, the choice of this threshold is important. In the case of \astrobj{$\beta$~Vir}, 
we iterated the process until all peaks with an amplitude higher than 3\,$\sigma$ in the amplitude
spectrum were removed (see Fig.~\ref{clean}). Peaks with amplitudes below the 3$\sigma$ threshold were 
not considered since they
were too strongly influenced by noise and by interactions between noise and daily aliases.
This threshold, which ensures that the selected peaks have
only a small
chance to be due to noise, gave a total of twenty-three frequencies (see Table~\ref{tab:identif}).
Because of the daily alias of 11.57\,$\mu$Hz introduced by the monosite observations, 
we cannot know a priori whether the frequency
selected by the algorithm is the right one or an alias. We thus considered
that the frequencies could be shifted by $\pm 11.57$ $\mu$Hz, and made echelle diagrams
for different large spacings near 72\,$\mu$Hz until every frequency could be identified as
an $\ell=0$, $\ell=1$ or $\ell=2$ mode. 
In this way, we found an averaged large spacing of 72.1\,$\mu$Hz.   

To investigate how many peaks are expected to be due to noise, we have conducted simulations in which we analyzed 
noise spectra
containing no signal. 
For this purpose, a velocity time series is built, using the observational time sampling
and radial velocities randomly drawn by assuming a Gaussian noise (Monte--Carlo
simulations).
The amplitude spectrum of this series is then calculated and peaks with
amplitude greater than 3,
4 and 5\,$\sigma$ are counted; note that a peak and its aliases are only counted once. 
The whole procedure is repeated 1000 times to ensure the
stability of the results. In this way, we find that the number of peaks due to noise with an amplitude
larger than 3\,$\sigma$ is 4.9\,$\pm$\,4.3 in the range 0.7--2.4~mHz, 
for 4\,$\sigma$, the number of peaks due to noise varies between 0 and 3 
with a mean value of 0.0 and a standard deviation of 0.5. 
No peaks due to noise are expected with an amplitude larger than 
5\,$\sigma$.

\begin{figure}
\resizebox{\hsize}{!}{\includegraphics{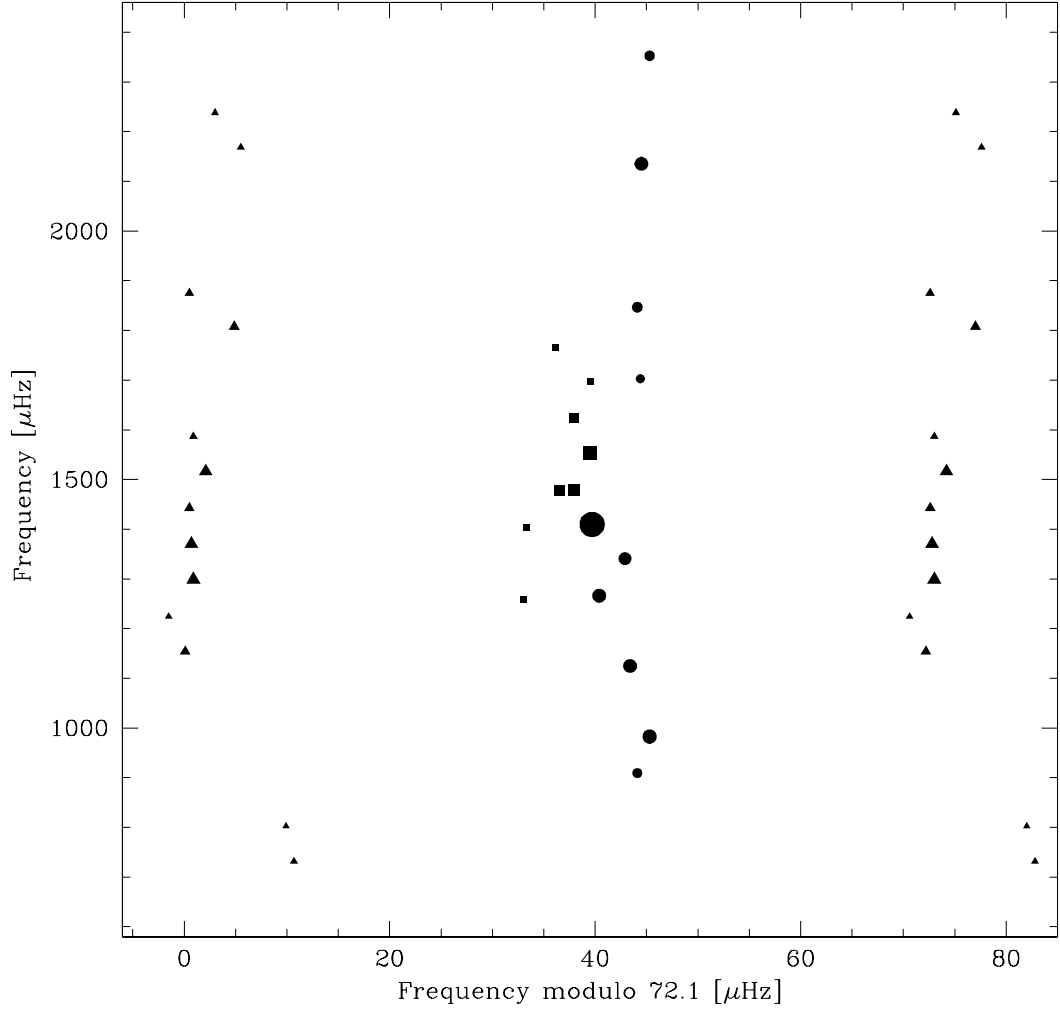}}
\caption{Echelle diagram of identified modes with a
large separation of 72.1\,$\mu$Hz. The
modes $\ell$=2 ($\blacksquare$), $\ell$=0 ({\Large $\bullet$}),
and $\ell$=1 ($\blacktriangle$) 
are represented with a size proportional to
their amplitude.}
\label{fig:ed}
\end{figure}

\begin{table}
\caption{Oscillation frequencies (in $\mu$Hz) of $\beta$~Vir. The frequency resolution
of the time series is 1.14\,$\mu$Hz.}
\begin{center}
\begin{tabular}{lccc}
\hline
\hline
 & $\ell$ = 0 & $\ell$ = 1 & $\ell$ = 2 \\
\hline
n = 8 &  & 731.7 &  \\
n = 9 &  & 803.0 &  \\
n = 10 &  &  &	  \\
n = 11 & 909.3 &  &    \\
n = 12 & 982.6 &  &        \\
n = 13 &  &  &  \\
n = 14 & 1124.9 & 1153.7 &    \\
n = 15 &  & 1224.2 & 1258.7 \\
n = 16 & 1266.1 & 1298.7 &	   \\
n = 17 & 1340.7 & 1370.6 & 1403.2   \\
n = 18 & 1409.6 & 1442.5 & 1478.5 / 1479.9  \\
n = 19 &  & 1516.2 & 1553.6	\\
n = 20 &  & 1587.1 & 1624.1\\
n = 21 &  & & 1697.8	\\
n = 22 & 1702.7 & & 1766.5\\
n = 23 &  &1807.3 &	\\
n = 24 & 1846.6 &1875.2 &	\\
n = 25 &  & &	\\
n = 26 &  & &	\\
n = 27 &  & &	\\
n = 28 & 2135.4 & 2168.4&	\\
n = 29 &  & 2238.2&	\\
n = 30 &  & &	\\
n = 31 & 2352.5 & &	\\
\hline
$\Delta\nu_{\ell}$ & 72.18 (0.09) & 71.91 (0.15)& 72.82 (0.30)\\
\hline
\end{tabular}\\
\end{center}
\label{tab:freq}
\end{table}
The echelle diagram showing the thirty-one identified modes is shown in Fig.~\ref{fig:ed}. 
The frequencies of the modes are shown in Fig.~\ref{tfiden} and are given in Table~\ref{tab:freq}, 
with radial order of each oscillation mode deduced from the asymptotic relation 
(see Eq.~1) assuming that the parameter $\epsilon$ is near the solar value 
($\epsilon_{\odot} \sim 1.5$).
The average large
spacing and the parameters $D_0$ and $\epsilon$ are thus deduced from a least-squares fit of this equation
with the frequencies of Table~\ref{tab:freq}. The average small spacing is here defined as $D_0 \times 6$ (the
perfect asymptotic case, with constant large and small separations):
\begin{displaymath}
\Delta\nu_{0}\,=\,72.07\,\pm\,0.10\;\mu Hz,\
\epsilon\,=\,1.59\,\pm\,0.03\;,
\end{displaymath}
\begin{displaymath}
D_0\,=\,1.05\,\pm\,0.24\;\mu Hz,\
\langle \delta\nu_{02} \rangle\,=\,6.28\,\pm\,1.42\;\mu Hz.
\end{displaymath}

The small spacing appears
to decrease from 7.4 to 4.9\,$\mu$Hz between 1260 and 1700\,$\mu$Hz.
\astrobj{$\beta$~Vir} has a projected rotational 
velocity of about $v$\,$\sin$\,$i$\,$\sim$\,4.3\,km\,s$^{-1}$ \cite{gl}.
We can estimate its radius to $R$\,=\,1.66\,$R_{\odot}$ from $L$\,=\,3.51\,$L_{\odot}$ (Hipparcos data) and $T_{\rm
eff}$\,=\,6140\,\degr\,K \cite{gray}. Assuming an uniform rotation, we determine the corresponding splitting of the modes
to be at least 0.56\,$\mu$Hz (for $\sin$\,$i$\,=\,1). This splitting can increase the uncertainty of the mode frequencies.
$\ell = 1$ modes can thus be shifted by at least $\pm$0.56\,$\mu$Hz, this effect is worse on
$\ell = 2$ modes (twice this value). Indeed, the dispersions of $\ell = 2$ and $\ell = 1$ modes are greater 
than for $\ell = 0$ modes (see Fig.~\ref{fig:ed} and
Table~\ref{tab:freq} with the larger errors on $\Delta\nu_{2}$ and $\Delta\nu_{1}$).
Moreover, two high $\ell = 2$ modes were determined with a same radial order (1478.5 and 1479.9\,$\mu$Hz), this could be
explained by such a splitting, implying an angle $i$ near 53\degr, however, with a high uncertainty.

Both $\ell=1$ modes at 731.7 and 803.0\,$\mu$Hz could be shifted by 11.57\,$\mu$Hz. They were fixed to their value in order to follow
the slight curvature present for $\ell=0$ modes at low frequency. They have to be taken with some caution.

\subsection{Oscillation amplitudes}
\label{sec amp}
Concerning the amplitudes of the modes, theoretical computations predict oscillation 
amplitudes near  50\,cm\,s$^{-1}$ for a 1.25 $M_{\odot}$ star 
like \astrobj{$\beta$ Vir}, with mode lifetimes of the order of several days to several tenth of days.
\cite{ho99}. The amplitudes of the highest modes, 
in the range 32--46\,cm\,s$^{-1}$, are then slightly lower than expected. 
The observations indicate that oscillation amplitudes
are typically 1.5--2 times solar. 
This disagreement can be partly explained by the lifetimes of the modes. 
Indeed, the oscillation modes have finite lifetimes,
because they are continuously damped. Thus, if the star is observed during a time longer than 
the lifetimes of the modes, the signal is weakened due to
the damping of the modes and to their re--excitation with a random phase.

\section{Conclusion}
	      
The radial velocity measurements of \astrobj{$\beta$~Vir}, obtained over 11 nights, show a significant 
excess in the power spectrum between 0.7--2.4\,mHz, 
centered around 1.4~mHz, with a maximal peak amplitude of 45\,cm\,s$^{-1}$,
revealing solar--like oscillations.
Moreover, we presented the identification of thirty-one individual frequencies. Note that this identification is a little complicated
by the presence of rotational splitting, as \astrobj{$\beta$~Vir} rotates rapidly enough to see this effect.
The signal-to-noise of our data is unfortunately not high enough to unambiguously determine this splitting.
The oscillation modes present an average large spacing of 72.1\,$\mu$Hz, and the small spacing appears
to decrease from 7.4 to 4.9\,$\mu$Hz between 1260 and 1700\,$\mu$Hz, with an average of 6.3\,$\mu$Hz. As the large spacing is related to $\sqrt{M/R^3}$
\cite{kb95},
the mass of \astrobj{$\beta$~Vir} should be close to 1.3\,M$_{\odot}$. Detailed theoretical models of \astrobj{$\beta$~Vir}
will be reported in a subsequent paper.

\section*{Acknowledgements}
This work was partly supported by the Swiss National Science Foundation.

\end{document}